\documentclass[twocolumn,showpacs,preprintnumbers,amsmath,amssymb,aps]{revtex4}
\usepackage{bm}
\usepackage{latexsym}
\usepackage{eufrak}
\usepackage[latin1]{inputenc}
\usepackage{latexsym}
\usepackage{amsmath}
\usepackage{epsfig}
\usepackage{ifpdf}
\usepackage{graphicx}
\usepackage{dcolumn}
\usepackage{amsmath}
\usepackage{amsfonts}
\usepackage{amssymb}
\begin{document}
\title{The Universe Dynamics from Topological Considerations}
\author{Miguel A. Garc\'{\i}a-Aspeitia\footnote{Part of the Instituto Avanzado de Cosmolog\'ia (IAC) collaboration http://www.iac.edu.mx/}} 
\email{agarca@fis.cinvestav.mx}
\author{Tonatiuh Matos$^*$}
\email{tmatos@fis.cinvestav.mx}
\affiliation{Departamento de F\'{\i}sica, Centro de Investigaci\'on y de Estudios Avanzados del I.P.N.
Apdo. Post. 14-740 07000, D.F., M\'exico}

\begin{abstract}
We explore the possibility that the dynamics of the universe can be reproduced choosing appropriately the initial global topology of the Universe. In this work we start with two concentric spherical three-dimensional branes $S^3$, with radius $a_1<a_2$ immersed in a five-dimensional space-time. The novel feature of this model is that in the interior brane there exist only spin-zero fundamental fields (scalar fields), while in the  exterior one there exist only spin-one fundamental interactions. As usual, the bulk of the universe is dominated by gravitational interactions. In this model, like in the Ekpyrotic one, the Big Bang is consequence of the collision of the branes and causes the existence of the particles predicted by the standard model in the exterior brane (our universe). The scalar fields on the interior brane interact with the spin-one fields on the exterior one only through gravitation, they induce the effect of Scalar Field Dark Matter with an ultra-light mass on the exterior one.
We discuss two different regimes where the energy density and the brane tension are compared, with the aim to obtain the observed dynamics of the universe after the collision of the branes. 
\end{abstract}

\keywords{Extra Dimension, dark matter theory, dark energy theory.}
\draft
\pacs{04.50.+h }
\date{\today}
\maketitle

\section{Introduction.}
In the last part of the twentieth century, cosmology became a
precision science getting observational data with an accuracy comparable with
the data obtained through the standard model. The last data arising
from observations confirm that our universe is accelerating due to the
existence of some unknown type of energy and also requires the existence of
an unknown field that permeates the universe and dictates the formation and
evolution of the structures at large scales. Both dark components of the
universe are not predicted by the standard model or by the general theory of
relativity either. This opens the possibility to extend these theories to limits beyond the current ones and to formulate new paradigms that
predict new physics, under the condition that in the appropriated limits these new theories reproduce the current standard observations.
So far the best accepted candidate to be the dark energy of the universe is the cosmological constant $\Lambda$. It is well-known that observations of the Cosmic Microwave Background Radiation (CMB) and galaxies surveys fit very well with a small value of $\Lambda$.  The main problem is to relate $\Lambda$ with some physical phenomena. There are many interpretations for doing so, for example, the most simple one is that the Einstein equations have two free constants, the gravitational one $G$ and the cosmological constant $\Lambda$. Here the cosmological constant contains only a geometrical interpretation. We fix $\Lambda$ in the same way as we have fixed $G$, namely, using cosmological observations. The problem here is that we have now a theory with two coupling constants, giving rise to the interpretation that the Einstein theory could not be the most fundamental theory for gravitation that can exist. Other example is to relate the cosmological constant with the vacuum energy of the universe. Here the problem is that the values derived from any particle physics models is many orders of magnitude too big (or vanishes in some SUSY cases) compared with the one obtained using astronomical observations \cite{2sean}. From this interpretation follows an incompatibility between the theory of general relativity and the particle physics models that we have to face on at this moment. There are some other proposals where the space-time contains subspaces embedded in a higher dimensional one, for example the branes theory
 \cite{7RS}, the DGP model \cite{4DGP}, Ekpyrotic \cite{Ovrut}, \cite{8Rasanen} or cyclic universe \cite{Turok}, \cite{6MacFadden} in which it is proposed the
existence of a four dimensional manifold embedded in a five dimensional
bulk. It is important to mention that in the last two models \cite{Ovrut}, \cite{8Rasanen} and \cite{Turok}, \cite{6MacFadden} the
branes move through the bulk and collide giving origin to the Big Bang.
The main goal of this work is to explore the possibility that the dynamics of the universe follows from its initial global topology. For doing so there are different ways, but the idea here is the following:
\begin{enumerate}
 \item We start with two three-dimensional branes embedded in a 5-dimensional space-time, where the five dimensional Einstein equations with cosmological constant are fulfilled. For facility we will suppose that the branes are three dimensional concentric spheres $S^3$ (FIG. \ref{fig:1}) but this result can be generalized to the other two homogeneous topologies $R^3$ and $H^3$.
 The initial conditions are such that the interior brane has the scale factor ${a}_{1}$ and the exterior brane (where we live) has the scale factor ${a}_{2}$, with ${a}_{1}<{a}_{2}$ (FIG. \ref{fig:1}). For simplicity, each brane have a common center.
 
\begin{figure}[htp]
\centering
\includegraphics[scale=0.3]{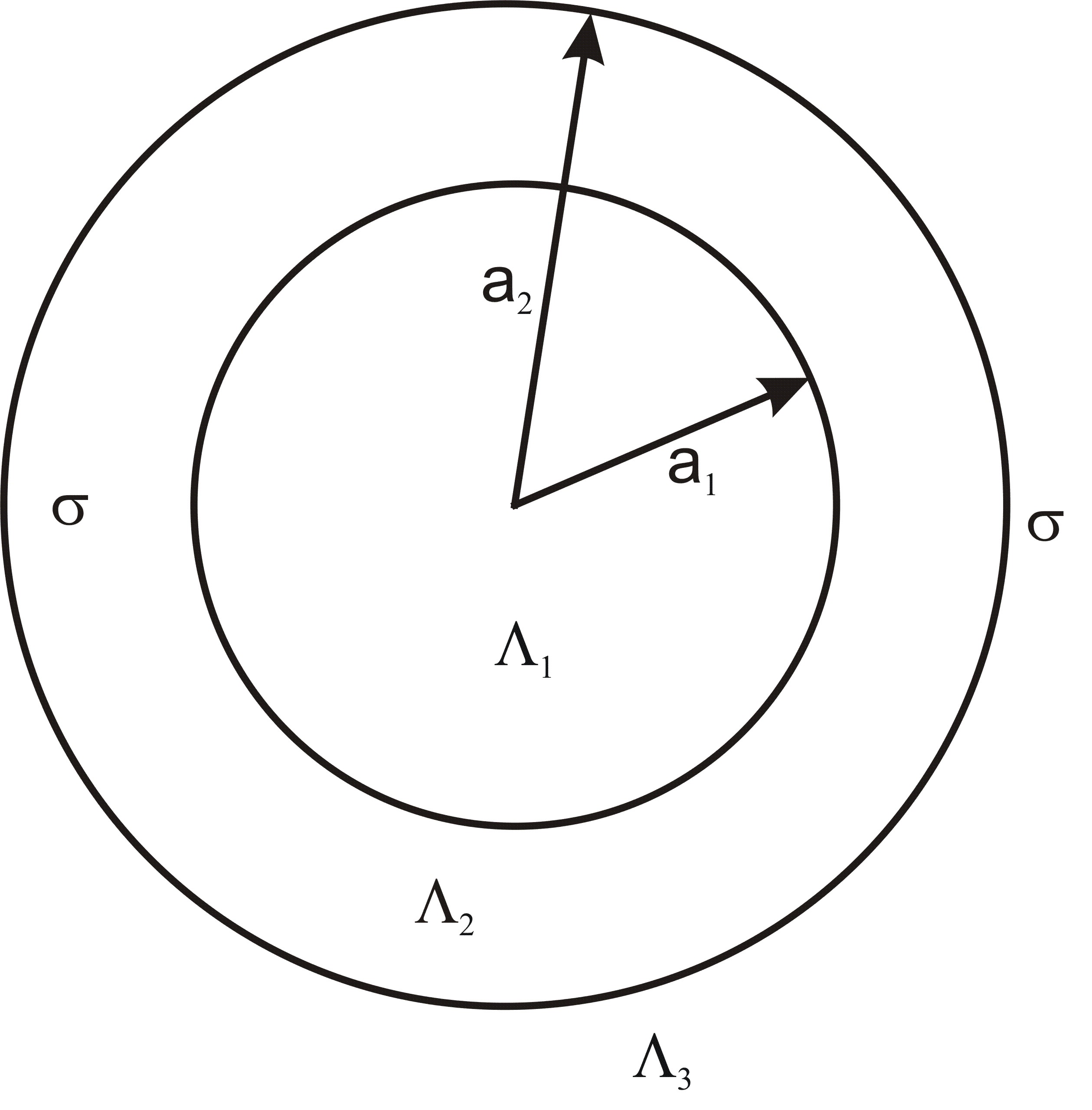}
\caption{Schematic representation of the interior and exterior branes and the different values of the 5D cosmological constant $\Lambda_{i}$ $i=1,2,3$ and the brane tensions $\sigma$.}\label{fig:1}
\end{figure}

\item We assume that matter is confined on the branes, the matter content of the interior brane is a self interacting scalar field (spin zero field) and of the exterior brane is the one of the standard model interactions (spin one fields).
\end{enumerate}

With these hypothesis we show that
\begin{enumerate}
 \item The scalar field on the interior brane acts as an inflaton, expanding this brane and provoking the collision with the exterior one, which was in thermal equilibrium before the collision. It follows a big bang scenario like in the Ekpyrotic models.
\item It is possible to choose the free parameters of the model such that after the collision, both branes expand together, (see equations (\ref{34}) and (\ref{35})). 
\item The fluctuations induced by the collision in the scalar field provoke the potential wells that give rise to the structure formation in the exterior brane, acting as the seeds for the galaxies formation on the brane where we live, via gravitational interaction. After the collision the physics of the exterior brane is close related with the Ekpyrotic and Cyclic models \cite{Turok}, \cite{Shiromizu}.
\end{enumerate}

The main difference of this toy model with previous ones is the hypothesis of the initial conditions and the matter content of the branes. For example, after the collision the exterior brane behaves in a very similar way as in the model presented in \cite{13liddleluis}. The problem of this last model is the reheating period, where it was not very natural to reheat the Universe. In the model presented here the reheating epoch is not necessary because the heat of the Universe is consequence of the collision of the branes, as in the Ekpyrotic model \cite{Turok}, \cite{Shiromizu}. 

It is important to mention that we let some open questions and use the results of previous works (see for example \cite{Shiromizu}, \cite{Turok}, \cite{8Rasanen}, \cite{6MacFadden}, \cite{Niz}), the main goal of this work is to show that there exist the possibility that the initial topology of the Universe could reproduce its observed dynamics. 

This work is organized as follows. In section \ref{sec_II} we present the model in detail and write the main field equations in section \ref{sec_III}. In section \ref{sec_IV} we present the different possible scenarios and regimes of the model. In section \ref{sec_V} we present the dynamics of the universe with this model and in section \ref{sec_VI} we give some conclusions and remarks.


\section{The Model.}\label{sec_II}
In this section we write the most important feature of the model in a general way. Suppose two concentric 3-dimensional branes embedded in a
5-dimensional bulk, (we use natural units $c=\hslash=1$). The shape of the action to model this physical structure is given by

\begin{eqnarray}
S&=&\int {dX^{5}}\sqrt{-{ g_{(5)}}}\,{ m}_{(5)}^{3}\left( {R_{(5)}+\Lambda }\right)\nonumber\\&&-\sum_{\pm }\int_{\pm }{dx^{4}}\sqrt{{ -g}^{\pm }}\left( { 2m}_{(5)}^{3}{ K}^{\pm }{ +\pounds}^{\pm }\right),  \label{1}
\end{eqnarray}
being $\pm $ the exterior or interior region of the brane respectively, $g_{(5)}$ is the determinant of the five-dimensional (5D) metric and $g$ the determinant of the four-dimensional (4D) one, ${ m}_{(5)}$ is the 5D Planck mass, $R_{(5)}$ is the 5D scalar curvature, $K$ is the extrinsic curvature and $\Lambda$ is the 5D cosmological constant, $\pounds^{+}$ is the Lagrangian of the fields content in the exterior brane (spin zero) and $\pounds^{-}$ is the Lagrangian of the fields content in the interior brane (spin one).
The bulk can be described with the natural coordinates in the form \cite{3ida}

\begin{eqnarray}
 { ds}_{(5)}^{2} &=&-A(a)_{\pm }dt_{\pm}^{2}{ +}\frac{{ 1}}{{ A}({ a})_{\pm }}{ da}^{2}\nonumber \\&+&a^{2}
\left[ { d\chi }^{2}+\sin^{2}(\chi)\left( { d\theta }^{2}+\sin ^{2}{ \theta d\varphi }^{2}\right) \right]. \label{2}
\end{eqnarray}
Now we solve the 5D Einstein equations ${ R_{(5)}}_{A B } = \frac{2}{3}\Lambda ^{\pm }{g_{(5)}}_{A B }$ with $(A,B=1,\cdots,5)$ for the most general $ A(a)_{\pm }$ function, we obtain

\begin{equation}
{ A(a)}_{\pm }=1-\frac{{ 2M}^{\pm }}{{ m}_{(5)}^{{ 3}}{ a}^{2}}{ -}\frac{{ \Lambda }^{\pm }}{{ 6}}{ a}^{2}, \label{3}
\end{equation}
where $\frac{2M^{\pm }}{m_{(5)}^{3}}$ is the mass parameter of the AdS-S$_{(5)}$ black hole \cite{6MacFadden} and $\frac{\Lambda ^{\pm }}{6}$ is the 5D cosmological constant in the bulk.

On the other hand, if we consider the location of the two 3-brane by $t=t(\tau )$, $a=a(\tau )$ parametrized by the time $\tau $ in the brane, we have
the induced four dimensional metric given by

\begin{equation}
ds^{2}_{(4)i}{ =-d\tau }^{2} +a^{2}_{i}(\tau )\left[{ d\chi }^{2}+\sin^{2}(\chi )\left( { d\theta }^{2}+\sin ^{2}( \theta)\, d\varphi^{2}\right) \right],  \label{4}
\end{equation}
where the subindex $i=1,2$ denote the two branes, having in mind that $a_{1}<a_{2}$. The functions $t(\tau )$ and $a(\tau )$ are then constrained by

\begin{equation}
{ u}^{\nu }{ u}_{\nu }=g_{\mu\nu}{ u}^{\mu }{ u}^{\nu } =-A(a_{i}){\dot t}^{2}+A(a_{i})^{-1}\dot a_{i}^{2} =-1,  \label{5}
\end{equation}
being ${ u}^{\mu }=\left( \dot t,\dot a_{i},0,0\right) $ the brane velocity. The dots represent here differentiation with respect to $\tau $. We define the unit normal vector on the two branes $n^{\pm}_\mu$ such that ${ n}_{\mu }^{\pm}{ u}^{\mu }=0$ and ${ n}_{\mu }^{\pm}{ n}^{\pm\,\mu} =1$, their components are given by

\begin{equation}
{ n}_{\mu }^{\pm }=\pm \left(\dot a_{i},-{ A}({ a_{i}})^{-1}({ A}({ a_{i}})+{\dot a_{i}}^{2})^{\frac{1}{2}},0,0\right).
\label{6}
\end{equation}
Following \cite{Shiromizu}, \cite{Maeda}, \cite{1cordero} we can construct the extrinsic
curvature using the equation

\begin{equation}
{ K}_{\mu \nu }{ =-g}_{\mu }^{\lambda }{ \nabla }_{\lambda}{ n}_{\nu }.  \label{7}
\end{equation}
The no null components of ${ K}_{\mu \nu }$ can be written as

\begin{equation}{ K}_{tt}^{\pm }=-\frac{\left({\ddot  a_{i}}+\frac{1}{2}\frac{\partial A(a_{i})_{\pm }}{\partial a_{i}}\right) }{\left( {\dot a_{i}}^{2} +A({ a_{i}})_{\pm }\right) ^{\frac{1}{2}}}, \label{8}
\end{equation}

\begin{equation}
{ K}_{\chi }^{\pm \chi }{ =K}_{\theta }^{\pm \theta }=K_{\varphi }^{\pm \varphi }=\frac{\left({\dot a_{i}}^{2}+A( a_{i})_{\pm }\right) ^{\frac{1}{2}}}{{ a_{i}}}. \label{9}
\end{equation}
With the equations of motion of the brane, it follows \cite{1cordero}, \cite{Shiromizu}, \cite{Maeda}

\begin{equation}
\left[ { K}\right] { g}_{\mu \nu }{ -[K}_{\mu \nu }{ ]=k}_{(5)}^{2}{ T}_{\mu \nu }{ ,}  \label{10}
\end{equation}

\begin{equation}{ \nabla }_{{ \nu }}{ T}_{{ \mu }}^{\nu }=-\left[ { T}_{{ \mu n}}\right]{,} \label{11}
\end{equation}
where $k_{ (5)}^{2} =\frac{8\pi }{m_{(5)}^{3}}$, $\left[ { K}_{\mu \nu }\right] =K_{\mu \nu }^{+}{ -K}_{\mu \nu }^{-}$ and $T_{\mu \nu}$ is the energy-momentum tensor.
Using equation \eqref{9} we find 

\begin{eqnarray}
\left( {\dot a_{i}}^{2}+A( a_{i})_{-}\right) ^{\frac{1}{2}}{ -}\left( {\dot a_{i}}^{2}+A(a_{i})_{+}\right) ^{\frac{1}{2}}=\frac{{ a_{i}}}{{ 3}}{ k}_{{ (5)}}^{2}{ \rho_{i} }{,} \label{12}
\end{eqnarray}
where we are supposing that the matter content of the branes can be represented as perfect fluids with density $\rho_i$ and presion $P_i$ \cite{1cordero}, \cite{Shiromizu}, \cite{Maeda}. 

The equation \eqref{11} represents the energy-momentum conservation on the brane, given by

\begin{equation}
\frac{d}{dt}\left({\rho_{i}{a_{i}^{3}}}\right)+P_{i}\frac{d}{dt}\left({a_{i}^{3}}\right)=0. \label{13}
\end{equation}
We will use the equation \eqref{12} to give a physical interpretation of the model.


\section{The modified Friedmann equations.}\label{sec_III}

In this section we write the field equations of the model. First observe that we have to distinguish between three vacuum regions, the first one is located inside the interior brane, we call it region I, the second one is located between the branes, region II, and the third vacuum region surrounds the exterior brane, region III, (see FIG. \ref{fig:1}). Now consider the following analogy with an electromagnetic system. Imagine two charged conducting parallel plates with opposite charges, we know that in the region in between the plates, there exist an electric field induced by the charges on the plates, while in the surrounded region, the electric field vanishes. If we now suppose that the branes are the plates and the electric charge is the gravitational field, the analogy tell us that it should exist an extreme difference between the expectation value for the vacuum in the region in between the branes and zero in the exterior region, having in mind that in the linear regime, the electromagnetic and the gravitational fields behave in the same way.

Thus, for the region inside of the interior branes (region I), equation \eqref{3} becomes

\begin{equation}
A(a_{1})_{-}=1 -\frac{{ \Lambda }_{1}}{{ 6}}{ a}_{1}^{2}{ ,}  \label{14}
\end{equation}
since there does not exist a gravitational potential inside of the interior brane. But in the region in between the branes (region II), the equation \eqref{3} can be
written as

\begin{equation}
A(a_{1})_{+}{ =1}{ -}\frac{{ 2M}_{1}}{{ m}_{(5)}^{3}{ a}_{1}^{2}}{ -}\frac{{ \Lambda }_{2}}{{ 6}}{ a}_{1}^{2}{ ,}  \label{15}
\end{equation}
where ${ M}_{1}$ is the mass of the interior brane. If we substitute the equations \eqref{14} and \eqref{15} into equation \eqref{12}, we have

\begin{eqnarray}
&&\frac{\dot{{ a}}_{1}^{2}}{{ a}_{1}^{2}}{ +}\frac{{ 1}}{{ a}_{1}^{2}}{ =}\frac{{ k}_{{ (5)}}^{{ 4}}{ \rho }_{1}^{2}}{{ 36}}{ +}\frac{{ \Lambda }_{1}
{ +\Lambda }_{2}}{{ 12}}{ +}\frac{{ M}_{1}}{{ m}_{(5)}^{3}{ a}_{1}^{4}}\nonumber \\&+&\frac{({ \Lambda }_{2}{ -\Lambda }_{1})^{2}}{{ 16\rho }_{1}^{2}{ k}_{{ (5)}}^{{ 4}}}{ +}\frac{{ 9M}_{1}^{2}}{{ m}_{{ (5)}}^{{ 6}}{ k}_{({ 5)}}^{{ 4}}{ \rho }_{1}^{2}{ a}_{1}^{8}}{ +}\frac{{ 3M}_{1}({ \Lambda }_{2}
{ -\Lambda }_{1})}{{ 2m}_{(5)}^{3}{ a}_{1}^{4}{ k}_{{ (5)}}^{4}{ \rho }_{1}^{2}}. \label{16}
\end{eqnarray}
On the other hand, inside of the exterior brane (region II again) we have

\begin{equation}
A(a_{2})_{-}{ =1}{ -}\frac{{ 2M}_{1}}{{ m}_{(5)}^{3}{ a}_{2}^{2}}{ -}\frac{{ \Lambda }_{2}}{{ 6}}{ a}_{2}^{2}{ .}  \label{17}
\end{equation}
Finally, outside of the exterior brane (region III), the total mass is ${ M=M}_{1}
{ +M}_{2}$. Therefore we have

\begin{equation}
A(a_{2})_{+}{ =1}{ -}\frac{{ 2}({ M}_{1}{ +M}_{2})}{{ m}_{(5)}^{3}{ a}_{2}^{2}}{ -}\frac{{ \Lambda }_{3}}{{ 6}}{ a}_{2}^{2}.  \label{18}
\end{equation}
Again, we substitute the equations \eqref{17} and \eqref{18} into the equation \eqref{12}, to obtain

\begin{eqnarray}
&&\frac{\dot {{ a}}_{2}^{2}}{{ a}_{2}^{2}}{ +}\frac{{ 1}}{{ a}_{2}^{2}}{ =}\frac{{ k}_{{ (5)}}^{{4}}{ \rho }_{2}^{2}}{{ 36}}{ +}\frac{{ \Lambda }_{3}{ +\Lambda }
_{2}}{{ 12}}{ +}\frac{{ 2M}_{1}{ +M}_{2}}{{ m}_{({ 5})}^{3}{ a}_{2}^{4}}\nonumber \\&+&\frac{({ \Lambda }_{3}{ -\Lambda }_{2})^{2}}{{ 16\rho }_{2}^{2}{ k}_{(5)}^{4}}{ +}\frac{{ 9M}_{2}^{2}}{{ m}_{(5)}^{6}{ k}_{(5)}^{4}{ \rho }_{2}^{2}{ a}_{2}^{8}}{ +}\frac{{ 3M}_{2}({ \Lambda }_{3}{ -\Lambda }_{2})}{{ 2m_{(5)}^{3}a}_{2}^{4}{ k}_{(5)}^{4}{ \rho }_{2}^{2}}. \label{19}
\end{eqnarray}
Expressions \eqref{16} and \eqref{19} are the Friedman equations on the interior and exterior branes respectively. In both equations, \eqref{16}) and \eqref{19}, the third, fourth, fifth and sixth terms are present only if the assumption of $\mathbb{Z}_{2}$-symmetry is dropped out.
Both equations, \eqref{16} and \eqref{19}, are in agreement with Ida et. al \cite{3ida}, except for the third, fourth, fifth and sixth terms which are characteristic of the proposed model.

\section{The Scenarios}\label{sec_IV}

According to the brane world scenario, it is natural to assume that the matter component consists of the brane tension $\sigma $ and the ordinary fields ${ \rho }_{ sf,}$ and $\rho_m$, such that the brane densities are given by

\begin{equation}
\rho _{1}=\rho _{sf}+\sigma, \;\; P_{1}=p_{sf} -\sigma , \label{20}
\end{equation}

\begin{equation}
\rho_{ 2} =\rho_{m}+\sigma, \;\; P_{2} =p_{m} -\sigma,  \label{21}
\end{equation}
where ${ \rho }_{{ m}}$  and ${ p}_{m}$ denote the energy density and the pressure of the matter with spin one interactions before the collision. After the collision we can interpret ${ \rho }_{{ m}}$  and ${ p}_{m}$ as radiation in the early universe (high energy regime) and as baryons plus radiation plus neutrinos, etc., in the late universe (low energy regime). ${ \rho }_{{ sf}}$ and ${p}_{sf}$  denote the energy density and the pressure of the self interacting scalar field respectively and $\sigma$ is the constant  brane tension \cite{3ida}.
 
Now we assume two scenarios on the branes, the first one corresponds to very big  densities and the second one when they are very small in comparison with the tension $\sigma$, that is \textbf{${ \rho }_{i}{ \gg \sigma }$} and ${ \rho }_{i}{ \ll \sigma }$, $i=sf,\, m$.

\subsection{High Energy Limit in the Early Universe.} 

First we analyse the branes when ${ \rho }_{i}{ \gg \sigma }$ , $i=sf,\, m$. This scenario corresponds to the universe before the collision.
We assume the next fine tuning conditions on the interior brane $\frac{1}{6}{ \Lambda }_{1}=\frac{2}{3}{ \Lambda }_{{ (5)}}$, $\frac{1}{6}{ \Lambda }_{2}=-\frac{2}{3}{ \lambda }_{{ (5)}}$, with ${ k}_{{ (5)}}^{4}{ =36}\frac{\kappa _{{ (4)}}^{2}}{3}\frac{1}{2\sigma }$ where ${\lambda_{(5)} \sim (10}^{18}{ GeV)}^{4}$ is of the order of magnitude of the quantum fluctuations of the vacuum predicted by the standard model  \cite{2sean}. We replace the previous conditions and equation \eqref{20} in equation \eqref{16} to obtain

\begin{eqnarray}
\frac{\dot {{ a}}_{1}^{2}}{{ a}_{1}^{2}}&+&\frac{{ 1}}{{ a}_{1}^{2}} =\frac{{ \kappa }_{{ (4)}}^{2}}{{ 3}}{ \rho }_{{ sf}}\left( { 1+}\frac{{ \rho }_{{ sf}}}{{ 2\sigma }}\right)\nonumber\\ &+&\frac{\overset{{(4)}}{{ \Lambda }}_{{ 1}}}{3}+\frac{{ M}}{{ m}_{{ (5)}}^{{ 3}}{ a}_{{ 1}}^{{ 4}}}\left[ 1-\frac{\sigma (\Lambda
_{(5)}+\lambda _{(5)})}{2\kappa _{(4)}^{2}\sigma \rho _{sf}\left( 1+\frac{\rho_{sf}}{2\sigma }\right) +\sigma ^{2}\kappa _{(4)}^{2}}\right] \notag \\
&&+\frac{{ 3M}^{{ 2}}}{{ m}_{{ (5)}}^{{ 6}}{ a}_{{ 1}}^{{ 8}}{ \kappa }_{{ (4)}}^{2}}\frac{{ \sigma }}{{ 4\sigma \rho }_{s{ f}}\left( { 1+}
\frac{{ \rho }_{s{ f}}}{{ 2\sigma }}\right) {+2\sigma }^{{ 2}}}{,} \label{22}
\end{eqnarray}
with

\begin{eqnarray}
\overset{{ (4)}}{{ \Lambda }}_{{ 1}}&=&\frac{{ \kappa }_{{ (4)}}^{2}{ \sigma }}{2}{ +(\Lambda }_{{ (5)}}{ -\lambda }_{{ (5)}})\nonumber\\&+&\frac{{ 
\sigma (\lambda }_{{ (5)}}{ +\Lambda }_{{ (5)}}{ )}^{{ 2}}}{{4\kappa }_{{ (4)}}^{2}{ \sigma \rho }_{{ sf}}\left( { 1+}\frac{{ \rho }_{{ sf}}}{{ 2\sigma }}
\right) {+2\sigma }^{{ 2}}{ \kappa }_{{ (4)}}^{2}} .\label{23}
\end{eqnarray}
This is because in the high energy limit it follows that 
\[{ \rho }_{sf}^{2}\sim2\sigma \rho _{sf}\left( { 1}+\frac{\rho _{sf}}{2\sigma }\right) { +\sigma }^{2},\]
 For simplicity we assume the ansatz $M_{1}=-M_{2}=-M$ with non physical interpretation. Then, imposing ${ \rho }_{sf}{ \gg \sigma }$ 
 and proposing $\kappa _{({ 4})}^{2}{ \sigma \approx \lambda }_{{ (5)}}$  we obtain

\begin{equation}
\frac{\dot{{ a}}_{1}^{2}}{{ a}_{1}^{2}}{ +}\frac{{ 1}}{{ a}_{1}^{2}}{ \approx }\frac{{ \kappa }_{{ (4)}}^{{ 4}}}{{ 3}}\frac{{ \rho }_{sf}^{2}}{{ 2\lambda }
_{{ (5)}}}{ +}\frac{\overset{{ (4)}}{{ \Lambda }}_{{ 1}}}{{ 3}}{ +}\frac{{ M}}{{ m}_{(5)}^{3}{ a}_{1}^{4}}{ ,}\label{24}
\end{equation}
with

\begin{equation}
\overset{{ (4)}}{{ \Lambda }}_{{ 1}} =-\frac{{ \lambda }_{{ (5)}}}{{ 2}}.\label{25}
\end{equation}
In the same way, for the exterior brane we can set the next fine tuning condition $\frac{1}{6}{ \Lambda }_{3}=\frac{2}{3}{ \Lambda }_{{ (5)}}$. 
Again, if one replace the previous conditions and equation \eqref{21} in equation \eqref{19}, we obtain

\begin{eqnarray}
\frac{\dot{{a}}_{2}^{2}}{{a}_{2}^{2}}&+&\frac{1}{a_{2}^{2}} =\frac{\kappa _{(4)}^{2}}{3}\rho _{m}\left( 1+\frac{\rho _{m}}{2\sigma }\right)\nonumber\\&+&{\small \frac{\overset{(4)}{\Lambda }_{2}}{3}}+\frac{M}{m_{(5)}^{3}a_{2}^{4}}\left[ -1+\frac{\sigma (\Lambda_{(5)}+\lambda _{(5)})}{2\kappa _{(4)}^{2}\sigma \rho _{m}\left( 1+\frac{\rho_{m}}{2\sigma }\right) +\sigma ^{2}\kappa _{(4)}^{2}}\right]   \nonumber \\&&+\frac{3M^{2}}{m_{(5)}^{6}a_{2}^{8}\kappa _{(4)}^{2}}\frac{\sigma }{4\sigma\rho _{m}\left( 1+\frac{\rho _{m}}{2\sigma }\right) +2\sigma ^{2}}, \label{26}
\end{eqnarray}
 with

\begin{eqnarray}
\overset{{ (4)}}{{ \Lambda }}_{{ 2}}&=&\frac{{ \kappa }_{{ (4)}}^{2}{ \sigma }}{2}{ +(\Lambda }_{{ (5)}}{ -\lambda }_{{ (5)}})\nonumber\\&+&\frac{{ \sigma (\lambda }_{{ (5)}}{ +\Lambda }_{{ (5)}}{ )}^{{ 2}}}{{4\kappa }_{{ (4)}}^{2}{ \sigma \rho }_{{ m}}\left( { 1+}\frac{{ \rho }_{{ m}}}{{ 2\sigma }}\right) {+2\sigma }^{{ 2}}{ \kappa }_{{ (4)}}^{2}}. \label{27}
\end{eqnarray}
Again,  we have set the limit condition
\[{ \rho }_{m}^{2}\sim2\sigma \rho _{m}\left( { 1+}\frac{\rho _{m}}{2\sigma }\right) { +\sigma }^{2},\]
Thus, in the limit where ${ \rho }_{m}{ \gg \sigma }$ and $\kappa _{({ 4})}^{2}{ \sigma \approx \lambda }_{{ (5)}}$, we obtain

\begin{equation}
\frac{\dot{{ a}}_{2}^{2}}{{ a}_{2}^{2}}{ +}\frac{{ 1}}{{ a}_{2}^{2}}{ \approx }\frac{{ \kappa^{4} }_{{ (4)}}}{{ 3}}\frac{{ \rho }_{m}^{2}}{{ 2\lambda }_{{ (5)}}}
{ +}\frac{\overset{{ (4)}}{{ \Lambda }}_{{ 2}}}{{ 3}}{ -}\frac{{ M}}{{ m}_{({ 5})}^{3}{ a}_{2}^{4}}{ ,}\label{28}
\end{equation}
with

\begin{equation}
\overset{{ (4)}}{{ \Lambda }}_{{ 2}}{ =-}\frac{{ \lambda }_{{ (5)}}}{{ 2}}.\label{29}
\end{equation}
The relations \eqref{24} and \eqref{28} are the Friedmann equations for the early universe of the model. Observe that the density $\rho_{sf}$ appears quadratic in these equations, thus, the brane containing the scalar field inflates, even with a very small scalar field mass and collides with the exterior one, where the matter content is only of spin one interactions and the brane expands without inflation.

\subsection{Low Energy Limit in the Late Universe.}

In the same way we analyse the regime ${\rho }_{i}{ \ll \sigma }$, $i=sf,\, m$ for the late universe using the same
conditions as in the previous regime. We obtain the next two relations for the
branes 1 and 2. 

\begin{eqnarray}
\frac{\dot{{ a}}_{1}^{2}}{{ a}_{1}^{2}}&+&\frac{{ 1}}{{ a}_{1}^{2}} =\frac{{ \kappa }_{{ (4)}}^{2}}{{ 3}}{ \rho }_{{ sf}}\left( { 1+}\frac{{ \rho }
_{{ sf}}}{{ 2\sigma }}\right)\nonumber\\&+&\frac{\overset{{ (4)}}{{ \Lambda }}_{{ 1}}}{3}+\frac{{ M}}{{ m}_{{ (5)}}^{{ 3}}{ a}_{{ 1}}^{{ 4}}}\left[ 1-\frac{\sigma (\Lambda_{(5)}+\lambda _{(5)})}{2\kappa _{(4)}^{2}\sigma \rho _{sf}\left( 1+\frac{\rho_{sf}}{2\sigma }\right) +\sigma ^{2}\kappa _{(4)}^{2}}\right] \notag \\
&+&\frac{{ 3M}^{{ 2}}}{{ m}_{{ (5)}}^{{ 6}}{ a}_{{ 1}}^{{ 8}}{ \kappa }_{{ (4)}}^{2}}\frac{{ \sigma }}{{ 4\sigma \rho }_{{ sf}}\left( { 1+}\frac{{ \rho }_{{ sf}}}{{ 2\sigma }}\right) {+2\sigma }^{{ 2}}}{ ,} \label{30}
\end{eqnarray}

\begin{eqnarray}
\frac{\dot{{a}}_{2}^{2}}{{a}_{2}^{2}}&+&\frac{1}{a_{2}^{2}} =\frac{\kappa _{(4)}^{2}}{3}\rho _{m}\left( 1+\frac{\rho _{m}}{2\sigma }\right)+\frac{\overset{(4)}{\Lambda }_{2}}{3} \\&&+\frac{M}{m_{(5)}^{3}a_{2}^{4}}\left[ -1+\frac{\sigma (\Lambda_{(5)}+\lambda _{(5)})}{2\kappa _{(4)}^{2}\sigma \rho _{m}\left( 1+\frac{\rho_{m}}{2\sigma }\right) +\sigma ^{2}\kappa _{(4)}^{2}}\right]   \nonumber \\&&+\frac{3M^{2}}{m_{(5)}^{6}a_{2}^{8}\kappa _{(4)}^{2}}\frac{\sigma }{4\sigma\rho _{m}\left( 1+\frac{\rho _{m}}{2\sigma }\right) +2\sigma ^{2}}, \nonumber \label{31}
\end{eqnarray}
with

\begin{eqnarray}
\overset{{ (4)}}{{ \Lambda }}_{{ 1}}&=&\frac{{ \kappa }_{{ (4)}}^{2}{ \sigma }}{2}{ +(\Lambda }_{{ (5)}}{ -\lambda }_{{ (5)}})\nonumber\\&+&\frac{{\sigma (\lambda }_{{ (5)}}{ +\Lambda }_{{ (5)}}{ )}^{{ 2}}}{{4\kappa }_{{ (4)}}^{2}{ \sigma \rho }_{{ sf}}\left( { 1+}\frac{{ \rho }_{{ sf}}}{{ 2\sigma }}
\right) {+2\sigma }^{{ 2}}{ \kappa }_{{ (4)}}^{2}},\label{32}\nonumber\\
\\
\overset{{ (4)}}{{ \Lambda }}_{{ 2}}&=&\frac{{ \kappa }_{{ (4)}}^{2}{ \sigma }}{2}{ +(\Lambda }_{{ (5)}}{ -\lambda }_{{ (5)}}) \nonumber\\&+&\frac{{ \sigma (\lambda }_{{ (5)}}{ +\Lambda }_{{ (5)}}{ )}^{{ 2}}}{{4\kappa }_{{ (4)}}^{2}{ \sigma \rho }_{{ m}}\left( { 1+}\frac{{ \rho }_{{ m}}}{{ 2\sigma }}
\right) {+2\sigma }^{{ 2}}{ \kappa }_{{ (4)}}^{2}}.\label{33}\nonumber\\
\end{eqnarray}
In the same way as before, this is because ${ \rho }_{i}^{2}=2\sigma \rho _{i}\left( { 1+}\frac{\rho _{i}}{2\sigma }\right) { +\sigma }^{2},{ (i=sf,m)}$.
Then, imposing the condition ${ \rho }_{i}{ \ll \sigma }$ , $i=sf,\, m$ and again $\kappa _{({ 4})}^{2}{ \sigma \approx \lambda }_{{ (5)}}$, we obtain

\begin{eqnarray}
\frac{\dot{{ a}}_{1}^{2}}{{ a}_{1}^{2}}{ +}\frac{{ 1}}{{ a}_{1}^{2}}&{ \approx }&\frac{{ \kappa }_{{ (4)}}^{2}}{{ 3}}{ \rho }_{sf}{ +}\frac{\overset{{ (4)}}{
{ \Lambda }}_{{ 1}}}{{ 3}}{ +}\frac{{ 3M}^{2}}{{ 2m}_{(5)}^{6}{ \lambda }_{{ (5)}}{ a}_{1}^{8}}{ ,}\label{34}
\\
\frac{\dot{{ a}}_{2}^{2}}{{ a}_{2}^{2}}{ +}\frac{{ 1}}{{ a}_{2}^{2}}&{ \approx }&\frac{{ \kappa }_{{ (4)}}^{2}}{{ 3}}{ \rho }_{m}{ +}\frac{\overset{{ (4)}}{
{ \Lambda }}_{{ 2}}}{{ 3}}{ +}\frac{{ 3M}^{2}}{{ 2m}_{(5)}^{6}{ \lambda }_{{ (5)}}{ a}_{2}^{8}}, \label{35}
\end{eqnarray}
where now we find the remarkable result that  
\[
\overset{{ (4)}}{{ \Lambda }}_{{ 1,2}} =\Lambda _{(5)}.
\]
Thus, we can fix the cosmological constant  $\Lambda _{(5)}\sim (10^{-12}{ GeV)}^{4}$ such that $\overset{{ (4)}}{{ \Lambda }}_{{ 1,2}} $ have the value of the classical observed 4-dimensional cosmological constant \cite{2sean}. The relations \eqref{34} and \eqref{35} are the Friedmann equations for the late time universe and give the dynamics of the branes. From the similitude of equations \eqref{34} and \eqref{35} we observe that both branes could have a very similar dynamics in the late time universe. We conclude that we can fix the free parameter of the model in order that both branes expand together. 

\section{The Dynamics}\label{sec_V}

\subsection{The Inflation in a Spherical Geometry.}

Now we can follow the dynamics of the branes. The main idea of the model we are dealing with here is not to fix the initial conditions $a_1(0)= something$, $a_2(0)= something$ else, etc. Instead of that we set the initial topology of the model as follows. Suppose there exist two concentric $S^3$ branes, in the interior one with a very small radius lives a scalar field; in the exterior brane live the particles of spin one. The matter contained in the exterior brane has a state equation $p=(\gamma-1)\rho$, which density evolve as $\dot\rho+3\frac{\dot a}{a}\gamma\rho=0$ with the equation (\ref{13}). During the early universe the expansion of the brane follows equation \eqref{28}, thus the exterior brane evolves as

\begin{eqnarray}
 \dot a_2^2+1=\frac{\kappa_{(4)}^4}{3}\frac{\rho_0^2}{2\,a_2^{6\gamma-2}\lambda_{(5)}}+\frac{\Lambda_2}{3}\,a_2^2-\frac{M}{m_{(5)}^3\,a_2^2}, \label{36}
\end{eqnarray}
or, if we derive this expression with respect to $t$, we arrive at

\begin{eqnarray}
 2\,\ddot a_2&=& -(6\gamma-2)\frac{\kappa_{(4)}^4}{3}\frac{\rho_0^2}{2\,a_2^{6\gamma-1}\lambda_{(5)}}+2\frac{\Lambda_2}{3}\,a_2\nonumber\\&+&2\frac{M}{m_{(5)}^3\,a_2^3}.\label{37}
\end{eqnarray}
If during this period the exterior brane is dominated by matter, the first term goes like $\sim a^{-5}$, if this brane is dominated by radiation, this term goes like $\sim a^{-7}$. In any case, during this period the first term dominates over the other two, indicating that this brane starts with a very big negative acceleration but decreasing its acceleration very fast. 

On the other brane the situation is very different due to the fact that the interior brane is permeated with a scalar field, its evolution is governed by equation \eqref{24}. During the epoch when the density is enough big, such that the two last terms on the right hand side of equations (\ref{24}) can be neglected, in the slow-roll approximation this equation reduces to

\begin{equation}
H^{2}\simeq\left(\frac{8\pi}{3m_{(4)}^{2}}\right)V\left[1+\frac{V}{2\sigma}\right], \label{38}
\end{equation}

\begin{equation}
\dot\Phi\simeq-\frac{V\prime}{3H}. \label{39}
\end{equation}
where $V$ is the scalar field potential. Then, using the two slow-roll parameters 

\begin{eqnarray}
 \epsilon&\simeq&\frac{2}{\kappa^2_{(4)}}\left(\frac{V'}{V} \right) ^2\frac{1+V/\sigma}{(2+V/\sigma)^2},\label{4o}\\\eta&\equiv&\frac{m_{(4)}^2}{8\pi}\left(\frac{V^{\prime\prime}}{V}\right)\left[\frac{2\sigma}{2\sigma+V}\right], \label{41}
\end{eqnarray}
if, for example, the scalar field potential is an exponential one $V=V_0\exp(-\alpha\kappa_{(4)}\phi)$, $\epsilon$ is given by \cite{copeland}

\begin{equation}
 \epsilon\simeq\frac{2\,\alpha^2\sigma}{V}. \label{42}
\end{equation}
During this period $V/\sigma>>1$ \cite{lidsey}, thus $\epsilon<1$ and the brane inflates till 
$\epsilon\sim 1$, this is, till the scalar field potential reaches the value
$V_{end}\sim 2\,\alpha^2\sigma$. If the potential is $V=\frac{1}{2}m_{\phi}^2\,\phi^2$, the slow-roll parameter reads 
\[\epsilon\sim \frac{8}{\kappa^2_{(4)}}\frac{1}{\phi^2} \frac{1+V/\sigma}{(2+V/\sigma)^2}=\frac{m^2_{Pl}}{\pi\,\phi^2}\,\frac{1+V/\sigma}{(2+V/\sigma)^2},\]
that means that if $V>>\sigma$, inflation ends when the scalar field reaches the value $\phi^2\sim\frac{1}{\pi\,}\,\frac{\sigma}{V}m^2_{Pl}$. For both potentials, the interior brane inflates and collide with the exterior one, heating the branes very fast. The heating of the branes essentially depends on the interaction constant between the scalar field and the matter and, of course, on the speed of the collision. Inflation in the interior brane also grows the quantum fluctuations of the scalar field, in such a way that after inflation ends, the interior brane contains a spectrum of semiclassical potential wells which become the seeds for the structure formation in the exterior brane.
Between these two regimes, during the branes collision, the physics of the system is just like in the Ekpyrotic model \cite{8Rasanen}. During the collision, quantum gravitational effects due to the interactions of both space-times take place.
From this moment, both branes follow the dynamics of equations \eqref{34} and \eqref{35}, indicating that both branes expand with the same dynamics. The scalar field riches the minimum of its potential implying  that, from this period on, the scalar field potential has a behaviour like $\sim\phi^2$. The scalar field in the interior brane induces potential wells into the exterior one via gravitational interaction, which evolve as cold dark matter (see \cite{aldebaran}). 

\subsection{The Scalar Field as Dark Matter.}

As we have explained before, the scalar field confined on the interior brane produces potential wells along the whole brane, caused by the extreme growing of the scalar field fluctuations during inflation. On the exterior brane, matter feels this potential wells as an extra gravitational force, but it cannot detect the scalar field in any other way.  In order to fit cosmological and galactic observations, the scalar field mass measured in the exterior brane must be ultra-light, such that $m_{sfdm}\sim10^{-22}$eV. This ultra-light mass causes, of course, a problem of hierarchy, however this problem can be solved using the prescription of Randall-Sundrum \cite{7RS} as follows.

The scalar field part of the action on the interior brane reads

\begin{equation}
S_{interior}=\int d^{4}x\pounds _{sf}=\int d^{4}x\sqrt{-g}\left[ g^{\mu \nu }\phi_{,\mu }\phi _{,\nu }+m_{sf}^{2}\phi ^{2}\right]   \label{43}.
\end{equation}
Using normal Gaussian coordinates for the AdS metric $ds^{2}=e^{-2ky}\eta _{\mu \nu }dx^{\mu }dx^{\nu }+dy^{2}$, action (\ref{43}) translated into the exterior brane reads

\begin{equation}
S_{exterior}=\int d^{4}x\left[ \eta ^{\mu \nu }\phi _{,\mu }\phi _{,\nu }+m_{sfdm}^{2}\phi^{2}\right],  \label{44}
\end{equation}
with the identification $e^{\xi y}\phi \longrightarrow \phi $  and $e^{-\xi y}m_{sf}=m_{sfdm}$, where $\xi^{2}=\frac{\Lambda}{6M_{(5)}^{3}}$. As long as the two branes expand together, $y$ remains constant after the collision. 
Thus, with this prescription it is enough that $\xi y\backsim{82}$, for instance, with the scalar field mass  $m_{sf}\sim5\times 10^5 M_{(5)}$ \cite{MIke}. That means, we obtain the required ultra-light mass of the scalar field on the exterior brane through a scalar field on the interior one. Furthermore, this ultra-light mass sets a minimum distance between the branes.

With this mass, that means,  with only one free parameter, the scalar field on the exterior brane behaves exactly in the same way as Scalar Field Dark Matter (SFDM) \cite{l9} with the following important features:

The ultra-light scalar field mass ($m_{SFDM} \sim 10^{-22}$eV) fits:
\begin{enumerate}
\item The evolution of the cosmological densities \cite{aldebaran}.

\item  The rotation curves of galaxies \cite{harko} and the central density profile of LSB galaxies  \cite{argelia}, 

\item With this mass, the critical mass of collapse for a real scalar field is just $10^{12}\,M_{\odot}$, i.e., the one obeserved in galaxies  haloes \cite{miguel}.

\item The central density profile of the dark matter is flat \cite{argelia}. 
 
\item The scalar field has a natural cut off, thus the substructure in clusters of galaxies is avoided naturally. With a scalar field mass of $m_\phi\sim10^{-22}$eV the amount of substructure is compatible with the observed one \cite{further}.

\item SFDM forms galaxies earlier than the cold dark matter model, because they form Bose-Einstein Condensates at a critical temperature $T_c >> $TeV. So, if SFDM is right, we have to see big galaxies at big redshifts.
 
\end{enumerate}

\section{Concluding Remarks}\label{sec_VI}

The question we are facing on here is the following; is it possible that the universe has the dynamics we observe because of its global topology? This question makes sense if we start from the 4D general Einstein equations in vacuum, $R_{\mu\nu}-1/2\,g_{\mu\nu}R-\Lambda\, g_{\mu\nu}=0$. This equation is completely geometrical, there is no matter content in it. The Friedman equation for this case is
\[\frac{\dot a^2}{a^2}+\frac{k}{a^2}-\frac{\kappa^2}{3}\Lambda=0,\]
where $k =0,\pm 1$ for a flat, $S^3$ or $H^3$ topology respectively. Or, in a more convenient form for our goal here, we can write it as
\[\frac{1}{2}\dot a^2-\frac{\kappa^2}{6}\Lambda\,a^2=-\frac{1}{2}k,\]
which can be seen as a $(1/2)\dot a^2 + V=E$ dynamical equation. We observe that even when there is no matter, the space-time posses a dynamics. Even if there is no cosmological constant $\Lambda=0$, it is enough that $k\neq0$ to have a dynamical universe. In other words, it is sufficient that at the beginning, the global topology is fixed to have a specific dynamics.

In this work we consider a special global topology of the universe, namely, two concentric spherical branes and some content of matter with different interactions in each region of the universe: fundamental interactions of spin zero confined in the interior brane, fundamental interactions of spin one confined in the exterior brane (the brane that will lead to the standard model) and gravitation in the bulk. We find that the spin zero interactions inflate the interior brane, making this brane to collide with the exterior one due to the concentric topology, we assume the branes heat because of a small interactions constant between the spin zero and the spin one particles.

After the collision, both branes expand together as shown by equations \eqref{34} and \eqref{35}. The scalar field fluctuations build potential wells which are seen as SFDM in the exterior brane via the RS prescription, with the idea showed in the equations (\ref{43}), (\ref{44}).
However, in the exterior brane we expect that the collision provokes the existence of particles with fractional spin and spin one, giving the zoo of particles predicted by the standard model like in the Ekpyrotic or Cyclic models \cite{Turok}. This idea is very interesting, but so far there does not exist mathematical evidence to back it up. It is important to remark that during the collision it is impossible to analyse all the interactions with the classical theory of general relativity due to the quantum-gravitational effects and the interactions between the fields that permeates both branes.

Thus, in this model the universe expands as we see it, but the particles of the standard model can not interact with the scalar field particles because they are confined in the interior brane and the standard model ones are confined in the exterior one. The five-dimensional cosmological constant $\Lambda$ is pure geometrical and can be seen as a second free constant of this toy model.
The extreme difference between the expectation values in the interior and exterior branes can be explained using an analogy with electromagnetism. If the fields in the 5D bulk produced by the branes, behaves like a electromagnetic field, remember that for a weak field, linearised Einstein equations are analogous to the electromagnetic one, then there exist a huge expectation value inside of the region between the two branes which is associated with $\lambda$, while in the exterior region the expectation value is zero. 

The benefit of working with spherical branes is the natural acceleration predicted by this kind of models where the acceleration starts at redshift $z>0.3$ in agreement with the observations \cite{Knop}. It is important to see that the acceleration does not depend much on the particular matter content, only on the geometrical features \cite{Gogbe}.

In future works it is necessary to analyze the perturbations on the exterior and interior branes \cite{MIke}, we expect that the scalar field fluctuations  reproduce the large scale structure, CMB \cite{Ivan}, Sachs-Wolfe effect  \cite{10maartens} etc. In the same way it is necessary to generalize the RS prescription in a global point of view where there exist a natural expansion and acceleration of the branes.

\section{acknowledgement}
 We want to acknowledge the enlightening conversation with Roy
Maartens during his visit in Leon Guanajuato, M\'{e}xico and many helpful discussions with Ruben Cordero, Abdel P\'erez Lorenzana and Luis Ure\~{n}a. This work was partially supported by CONACyT M\'exico, under grants 49865-F, 216536/219673 and  I0101/131/07 C-234/07, Instituto Avanzado de Cosmologia (IAC) collaboration.


\begin{thebibliography}{99}

\bibitem{1cordero} Cordero Ruben and Vilenkin Alexander., \emph{et al}., Phys. Rev. D \textbf{65}, 083519. arXiv.hep-th/0107175

\bibitem{7RS} Randall Lisa and Sundrum Raman., \emph{et al}., Phys. Rev. Lett. \textbf{83} 3370-3373 (1999), arXiv:hep-ph/9905221v1

\bibitem{8Rasanen} R\"{a}s\"{a}nen S., Ph.D. thesis, (2002). arXiv:astro-ph/0208282v2 

\bibitem{miguel} Alcubierre Miguel, Siddhartha Guzm\'an F., Matos Tonatiuh, N\'u\~nez Dar\'{i}o, Ure\~na Luis A. and Wiederhold Petra. Class. Quant. Grav. \textbf{19} 5017-5024 (2002), gr-qc /0110102.

\bibitem{argelia} Bernal Argelia, Matos Tonatiuh and N\'u\~nez Dar\'{\i}o., Rev. Mex. A.A. \textbf{44}, 149-160 (2008), arXiv:astro-ph/0303455

\bibitem{2sean} Carroll Sean M. Living Reviews in Relativity, (1999), arXiv:astro-ph/0004075v2

\bibitem{11coley} Coley Alan A. Proceedings of ERE-99, arXiv:gr-qc/9910074v1

\bibitem{copeland} Copeland Edmund J, Liddle Andrew R. and Lidsey James E \emph{et al}., Phys. Rev. D \textbf{64}, 023539 (2002).
%

\bibitem{9Dodelson} Dodelson Scott. Academic Press, Copyright 2003, Elsevier

\bibitem{4DGP} Dvali G., Gabadadze G., Porrati M. \emph{et al}., Phys. Lett. B \textbf{485} 208-214, (2000), arXiv:hep-th/0005016v2

\bibitem{Gogbe} Gogberashvili Merab., \emph{et al} Phys. Lett. B \textbf{636} 147-149 (2006) arXiv:gr-qc/0511039v2

\bibitem{3ida} Ida D. Extra Large Dimensions, Classical Theories of Gravity  JHEP09 (2000) 014.

\bibitem{Ovrut} Khoury, Ovrut, Steinhardt, Turok., \emph{et al}., Phys. Rev. D \textbf{64}:123522 (2001)

\bibitem{Knop} Knop R. A, \emph{et al}., Astrophys. J. \textbf{598} 102 (2003) arXiv: astro-ph/0309368; 
A. G. Riess, \emph{et al}., Astrophys. J. \textbf{607} 665 (2004) 665 arXiv: astro-ph/0402512

\bibitem{l9} Matos Tonatiuh and Siddhartha Guzm\'an F. \emph{et al}., Class. Quant. Grav. \textbf{17}, L9-l16 (2000), arXiv: gr-qc /9810028

\bibitem{aldebaran} Matos Tonatiuh, Vazquez Alberto and Maga\~na J.A. \emph{et al} MNRAS., \textbf{389}, 13957 (2009) arXiv:0806.0683
 
\bibitem{further} Matos Tonatiuh and Ure\~na Luis A., \emph{et al}., Phys Rev. D \textbf{63}, 063506 (2001), arXiv:astro-ph/ 0006024

\bibitem{12Langlois} Langlois D., Maartens R., Sasaki M., Wands D. \emph{et al}., Phys. Rev. D \textbf{63}, 084009 (2001), arXiv:hep-th/0012044

\bibitem{13liddleluis} Liddle A. and Ure\~a-L\'opez L., \emph{et al }., Phys. Rev. Lett. \textbf{97}, 161301 (2006).

\bibitem{lidsey} Lidsey James, Matos Tonatiuh and Ure\~na Luis A. \emph{et al}., Phys Rev. D \textbf{66}, 023514 (2002), arXiv:astro-ph/ 0111292

\bibitem{10maartens} Maartens Roy., \emph{et al} Living Rev. Rel.7:7, (2004), arXiv:gr-qc/0312059v2

\bibitem{maartens} Maartens R., Wands D., Bassett B.A. and Heard I.P.C. \emph{et al} Phys. Rev. D \textbf{62}, 041302 (2000).
%

\bibitem{6MacFadden} Mac Fadden P., PhD thesis. arXiv:hep-th/0612008v2


\bibitem{Maeda} Maeda, Mizuno, Torii., \emph{et al}., Phys. Rev. D \textbf{68}, 024033 (2003)

\bibitem{Niz} Niz Quevedo Gustavo., PhD thesis, (2006)

\bibitem{5abdel} Lorenzana Abdel., \emph{et al}., J.Phys.Conf.Ser. \textbf{18} 224-269 (2005), arXiv:hep-ph/0503177v2]

\bibitem{Ivan} Rodriguez Montoya Ivan and Matos Chassin Tonatiuh., (2009) arXiv:0908.0054v1 [astro-ph.CO]

\bibitem{Shiromizu} Shiromizu, Maeda, Sasaki., \emph{et al} Phys. Rev. D \textbf{62} 024012 (2000)

\bibitem{Turok} Steinhardt, Turok., arXiv:hep-th/0111098

\bibitem{MIke} Miguel A. Garc\'ia Aspeitia, Tonatiuh Matos, Juan Maga\~na and Pablo A. Rodriguez., In preparation.

\bibitem{harko} C. G. Boehmer and T. Harko., \emph{et al}., JCAP \textbf{0706} 025 (2007), arXiv:astro-ph/0303455.

\end{thebibliography}
\end{document}